\documentclass[11pt]{article}
\usepackage{graphicx}
\usepackage{amssymb}

\font\sf=cmss10 at 9pt %san serif
\font\ssf=cmss10 at 9pt
\font\ssf=cmss10 at 6pt

\font\bb=msbm10 at 9pt
\font\bbs=msbm10 at 7pt

\font\bfs=cmb10 at 7pt

\font\rmss=cmr10 at 7pt

\font\cals=cmsy10 at 8pt

\def\0#1{\mbox{\rm #1}}
\def\1#1{\mbox{\bb #1}}
\def\2#1{\mbox{\bf #1}}
\def\3#1{{\mathcal #1}}
\def\4#1{\mbox{\cals#1}} 
\def\5#1{\mbox{\sf #1}} %%sans serif
\def\6#1{\mbox{\ssf #1}}
\def\7#1{\mbox{\bfs #1}}
\def\8#1{\mbox{\rmss #1}}
\def\9#1{\mbox{\bbs #1}}

\title{{\Huge \bf THE QUBITS OF QUNIVAC}\\
{\large For the proceedings of}\\
Digital Perspectives\\
{\large NSF, Arlington, VA, 2001}
}
\author {James Baugh,
David Ritz Finkelstein, Andrei Galiautdinov\\
Georgia Institute of Technology}
\begin{document}
\maketitle

\abstract{
We formulate a theory of quantum processes,
extend it to
a generic quantum cosmology, 
formulate a reversible quantum logic
for the Quantum Universe As  Computer, 
or Qunivac.
Qunivac 
has an orthogonal group of cosmic dimensionality.
It has a Clifford algebra of ``cosmonions,''
extending the quaternions to
a cosmological number of anticommuting units.
Its qubits
obey Clifford-Wilczek statistics and
are associated with unit 
cosmonions.
This makes it relatively easy to program the Dirac equation
on Qunivac in a Lorentz-invariant way.
Qunivac accommodates a field theory and a gauge theory.
Its gauge group is necessarily a quantum group.
}
\section{Overview}

Most of us at this symposium suppose 
that the universe is a computer.
If so,
then it seems likely that it is
a reversible quantum relativistic computer.

To carry quantum theory
that far from 
the spectroscopy laboratory
we must 
package it carefully,
especially since 
there is a virus
going around.

Then we must formulate a reversible quantum logic.
The familiar quantum logic is based on irreversible
filtration operations
combined by irreversible lattice operations
of $\cup$ and $\cap$.

Here
we express quantum theory in a 
way suitable for cosmology
and provide
an antiviral mantra (Part 1),
extend this quantum theory 
from laboratory systems to the cosmos,
imitating Laplace (Part 2),
and finally specialize to the quantum universe as
computer
with a useful correspondence principle (Part 3).
We formulate a reversible logic
and then a reversible quantum logic.
It is then straightforward to
program the Dirac equation and 
generic field
and gauge theories
in a clearly Lorentz-invariant way.

\part{\Huge Quantum systems have no state}

An ancient philosophical virus, 
already described by Francis Bacon,
is now at large in the quantum-physics community.
Most of us conferees carry it here today,
and we unwittingly 
infect our students as we were infected by our teachers.

This subtle virus 
leaves the mathematical part of  physics
untouched. 
The calculating engine of the infected theorist 
runs at full speed and precision.
Only the io (input-outtake) function is 
disturbed,
so that the end result is corrupted.
Most of those infected by this virus
have learned to compensate for it
by 
using one theory and 
speaking another.
At the same time,
some physicists 
express and use quantum theory 
self-consistently.

As a result, there are today two 
versions of quantum physics.
One is practiced widely and works,  but  is rarely professed.
The other is widely professed and never practiced,
not even by its adherents,
being inconsistent with actual practice.

The discord appeared in the early days of quantum theory.
Shortly after Heisenberg invented matrix mechanics,
based on 
processes or operations,
Schr\"odinger invented a wave mechanics,
an ontological theory.
This did not work at first, 
but it could be after-fitted
with {\em ad hoc}  rules 
that made it consistent with 
experiment and matrix mechanics.
The ontological theory
propagated more rapidly than the operational one,
perhaps because it is visualizable,
and now  has almost driven the processual 
one out of the classroom.

But the source of the infection is even older.
The virus resides in natural language.
Just as natural language conflicts with special relativity in its tense structure,
it conflicts with quantum physics in its
predication structure.
In both cases natural language assumes a non-existent now,
an ``is''  that actually isn't.
Bacon would call this ``is''  an idol of the tribe \cite{BACON}.
The virus is an ontology;  we may as well call it
an ontovirus.
Mathematical hubris,
the belief that mathematical symbols can
represent nature faithfully,  predisposes  one  to ontoviral infection.

In this part we juxtapose and align sample formulations of two classes 
of theories
that we call praxic and ontic respectively
\cite{FINKELSTEIN1996},
and which evolved from and generalize
matrix mechanics and wave mechanics.

\section{Praxic theory}

{\Large  Ideal  input and outtake
 processes
represented by vectors $|i\rangle$
and dual vectors $\langle o |$ 
have the transition probability 
\begin{equation} 
\label{eq:MALUS}
P= \cos^2 \theta := |\langle o |i\rangle|^2\/.
\end{equation}
}
$P$ is the square of the projection of the intake vector on the outtake vector.
Evidently complex phase factors in the two vectors
are ignorable.

\section{Ontic theory}

{\em During an ideal yes-or-no measurement 
a state represented
by a vector $|i\rangle$
changes 
to a state represented by a vector $|o\rangle$
with probability}
\begin{equation}
\label{eq:WIGNER}
P=\cos^2 \theta := |\langle o |i\rangle|^2\/.
\end{equation}

\section{The virus}

What a fiendish virus!
In their mathematical structures,
 (\ref{eq:MALUS}) and (\ref{eq:WIGNER})
are identical. 
There is no way to tell that one 
theory is sound, one unsound,
from within the mathematical theory.
We have to watch them in use to discover this.

In the praxic formulation the vectors represent
processes.
The ontic theory has mistaken them for 
states,
which then must undergo other processes.

The ontovirus is apparent
when \cite{NEUMANN}
speaks of two modes of intervention
${\bf 1}$ and ${\bf 2}$
into a quantum system,
representing  measurement and propagation.
The indication is that
there are three buildings, not two,
at accelerator facilities,
housing three modes of intervention,
${\bf 1, 2, 3}$,
not two,
 of
inflow, throughflow, and outflow;
or beam preparation,
target interaction,
and counting.
They are represented by the three factors
in the transition amplitude
$\langle o | T| i\rangle$.
They  exist in the simplest quantum experiment too.

The ontic
theory does not count the input process.
It miscounts because
it mistakes that process for the thing produced
and so counts as processes  
what are actually
intervals between processes.
A figure-ground reversal 
has occurred.

Three processes have two intervals;
so the ontist imagines a physical process $\bf 2$
 that transforms $|i\rangle$ into
$T|i\rangle$, and a process $\bf 1$  that converts 
$T|i\rangle$ into $\langle o|$.

No such things occur In the laboratory;
$|i\rangle$ does not transform into 
$T|i\rangle$ which changes into $|o\rangle$.
$|i\rangle$ is simply followed by $T$,
which is followed by $\langle o|$.

\section{Malus}

There is no way 
to tell whether the praxic or ontic theory
is right  from the mathematical theory.
The error is in the semantics.
The use of the theory is described incorrectly,

We must watch the working physicist using the theory
to describe the use correctly.
The earliest quantum experiment suffices.
Malus \cite{MALUS} considered  a photon that has
passed undeflected  through one crystal of Iceland spar
--- that is
the input process $|i\rangle$ ---
and is about to meet another --- 
the outtake process  $\langle o|$ .
(\ref{eq:MALUS}) is Malus' law
for the probability that the photon 
will again be undeflected.

A photon polarization in flight along an optical bench
--- say on the $z$ axis --- is postulated
by the ontic theory to have a state
$\psi(z,t)$ at time $t$,
a unit vector of two complex components,
with overall phase ignored.

It happens that there already is a familiar physical system 
that has such a state.
In the ontic theory, a photon polarization
is merely a particle moving on a sphere
with a special first-order dynamical equation;
except that unlike such a particle
it jumps in a certain probabilistic way
when 
we do what quantum physicists 
persist in calling
a measurement of the polarization
along some chosen direction.

This is a misnomer according to the ontic theory,
which claims that the process is actually a 
certain kind of kick of the particle state,
not a measurement at all.
We never do what the ontist
could honestly call observing the particle,
which is  to measure its state $\psi$
at some time.

\subsection{The test}

To see what $\psi$ actually is, then,
we watch a physicist in action
and note what she does;
not what she says,
of course,
since she may carry the ontovirus.

In this {\em Gedankenexperiment}  
we lead a trained quantum physicist
to the optical bench and ask her
 to estimate whether a certain photon
--- say the first after high noon ---
that has passed undeviated through the first crystal will pass
undeviated through the second crystal.

She knows not to make any further  
 measurement
on said photon
in flight  between the polarizers,
because that could change it and the outcome.
She might measure the angle $\theta$ between the input crystal
and the outtake crystal, and use Malus' law.
She might put a billion other
photons through a like process,
count the fraction that pass the test,
and use that fraction as the probability for the given photon.
But in any case her practice is not the astronomer's.
He can  look at the actual system to tell where it is going and whence it came.
She cannot.
She looks at the polarizers,
not the photon, for the polarizer angle;
and nothing she can do to the photon
will give her that information.

We should not call that angle the state of the photon
because, 
whether we call it a state or not,
it is not of the photon.

Transition probability and $|i\rangle$ and $\langle o|$
are features of the process, 
not of its product.
A $\psi$ does not evolve into a $\phi$
on the optical bench.
We choose  both freely 
when we set up the two crystals.
They are not the kind of things that evolve,
we just do them or not.
Having done them,
we can use Malus' Law to estimate the transition probability,
the probability of the {\bf 3} going conditional upon the {\bf 1} having gone.

It was not a mystery
why some took the input process  for its product
in a first formulation.
In classical physics,
the input process, the state, and the output process
of an allowed transition
all determine each other
for purposes of  prediction and retrodiction.
The classical observer could look at any of them
to determine the state.

The quantum physicist, however,
 does not have that choice,
but must observe and consider
all the processes,
which are almost independent of each other
in the allowed transitions.

In one breath the ontic theory loads 
the photon 
with an infinity of information
in its ``state,''
and in the next breath
denies that the photon can divulge one bit of that information in 
a measurement.
This is the kind of theory that our fathers warned us against.
It feigns a hypothesis.

The  mathematical problems of the quantum theory
all correspond exactly to problems of the ontic theory,
but the ontic theory is wrong for the quantum polarization.
The natural-language term  ``state of the system'' 
has a reserved meaning in physics.
We err if we call  data 
gotten by observing the state of Venus
``the state of Mars''  or ``the state of the astronomer.''
Everywhere else,
what  we call the state of a system 
is understood to be something that we can 
in principle learn  from the system itself
and use to predict
the system's future behavior.
Trying to override  this fact of natural language has
led to serious and costly distortions.

\subsection{Quanta have no state}

The situation was  clearly  formulated  by  
Bergmann \cite{BERGMANN}.
The concept of state is inappropriate for quanta.

Quantum theory is 
a theory of quantum processes.
It is no more a theory of a state
 than special relativity is a theory of a  present.
This is why Heisenberg  called his theory non-objective.
and why  Blatt and Weisskopf
refer to $\psi$'s as channels, not states \cite{BLATT}.
A $\psi$ describes the  process, 
not the product of the process.
There is no problem of ``collapse'' 
of the state in quantum theory
because there is no state to collapse.

In one breath the ontic theory loads 
the photon 
with an infinity of information
in its ``state,''
and in the next breath
denies that the photon can divulge one bit of that information in 
a measurement.
This is the kind of theory that our fathers warned us against.
It feigns a hypothesis.

The  mathematical problems of the quantum theory
all correspond exactly to problems of the ontic theory,
but the ontic theory is wrong for the quantum polarization.
The natural-language term  ``state of the system'' 
has a reserved meaning in physics.
We err if we call  data 
gotten by observing the state of Venus
``the state of Mars''  or ``the state of the astronomer.''

Everywhere else,
what  we call the state of a system 
is understood to be something that we can 
in principle learn  from the system itself
and use to predict
the system's future behavior.
Trying to override  this fact of natural language has
led to serious and costly distortions.

The discord between quantum practice and 
ontic principle has created unease in the most thoughtful of
the affected
physicists,
and 
the better the quantum theory 
works, the greater the unease.
The work devoted to resolving this unease has been immense:
the quantum potential, the many-world theory, 
decoherence theory,
consistent histories:
all 
manifestations  of  the ontovirus.

A $\psi$ is neither a quantum nor the state of one.
It is a process that can produce one.

We suggest the following mantra for those exposed to the ontovirus:

\begin{quotation}
$\psi$:  process, not object.
 \end{quotation}

\part{Quantum cosmology}

Quantum practitioners have never been troubled my
philosophical qualms when they set about making a 
quantum theory of the universe.
Every quantum field theory since Dirac's quantum
electrodynamics has been 
a 
quantum theory of the universe.

These theories are conspicuously non-operational.
No one in the universe
can prepare or register it all sharply.
Any physical experimenter is made of the very 
particles and fields of the theory
and so is in the field system,
not outside it experimenting on it.

At first Bohr objected strenuously to a quantum theory
of the universe for this reason, but later he 
withdrew his objections \cite{BOHR}.
It has always been understood,
at least tacitly,
that to correspond theoretical descriptions to those
of a physical experimenter,
one simply ignores the variables that
the experimenter ignores,
including those of the experimenter.

For those who think that a quantum system has a state,
however,
it is only one step to thinking that the universe has one.
This is the ontovirus on a cosmic scale.

The  quantum cosmology
of field theory
and the one we use here
corresponds 
--- in the sense of Bohr's correspondence principle ---
to
Laplace's classical cosmology
and is just as natural.
Both cosmologies  metaphorically deify the physicist.

Laplace invented a supreme astronomer 
who knows the state of the universe.
Correspondingly
we may imagine a supreme quantum Cosmic Experimenter (CE)
who
inputs the polarized cosmos with a grand $|I\rangle$ 
before our experiments
and outtakes it with an $\langle O|$ after our experiments.
We too need the CE
to formulate a cosmology as much as Laplace did,
but if we want to do quantum cosmology she has to be a quantum 
Experimenter.

These cosmologies are not operational.
They deal with metaphorical processes carried out by 
the metaphorical CE.
We extract  operational predictions
from them just as Laplace would.

We describe an actual experimenter as a subsystem of the cosmos
with its own algebra,
and  ignore or average over
the  degrees of freedom
of the cosmos that the actual experimenter ignores,
especially those of that very experimenter.

If the  CE
 were to have set us up before the fact to
do our little experiments
and were to read our notebooks after the fact,
then her readings would be consistent 
with ours \cite{NEUMANN}.

The extramundane CE
 is our metaphorical way of allowing 
for all possible experimenter/system interfaces
within one embracing theory.
The cosmic algebra is not operational but
contains operational algebras
as quotients.
Quantum theory
was already
the most relativistic theory we have.
Quantum cosmology relativizes it further.
It relativizes the experimenter.

Those affected by the ontovirus can function just as they do 
in laboratory applications.
Having  imagined a cosmic state-vector,
 they must imagine cosmic
observer  to collapse it by an observation at the end of time.

\part{Qunivac}

Now we describe the structure of the cosmic computer
from this extramundane viewpoint.
We begin with the algebra of the universe.

\section{Quantization is stabilization}

The standard model of the
elementary particles has several non-semisimple groups:
groups that are reducible but not decomposable.
So does Einstein's 
model of gravity.
All such theories are unstable
with respect to small
variations in their
structure tensors
\cite{SEGAL, INONU}.
This means that they are
singular limits of 
several deeper,
stabler theories
that  preserve all
the basic principles 
of quantum theory and
relativity,
at least asymptotically, and
are more unified.
One of these stabler theories
probably fits experiment better
than the present unstable theory
\cite{SEGAL}.
The unstable theory might be right, 
but this
has probability 0.

One of the deeper instabilities
--- not the deepest ---
is that of the 
differential calculus
and the space-time continuum
\cite{SEGAL}.
Systematic stabilization therefore
quantizes the differential calculus
and the space-time continuum.
Symmetry and stability arguments
lke those of \cite{SEGAL}
 led us to postulate 
an
elementary-process hypothesis
to replace and unify
the continuum hypothesis 
and the atomic hypothesis:

{\large \bf 
All physical processes
are composed of finitely many finite
elementary quantum processes.
\cite{FINKELSTEIN1969,WHEELER,FINKELSTEIN1996}}

We assert this for space-time translations and boosts
as well 
as particle creation and annihilation.
We assume that the elementary process
--- may we call it a pragmon for short? ---
lasts at least a minimum time $\Delta \tau$,
and 
transfers at most a maximum energy $\Delta \epsilon$.

\section{Reversible quantum logic}

We assemble qubits into  Qunivac
by imitating how one assembles bits into a classical
computer.

Classical computers are usually defined conceptually 
using set theory,
and set algebra is usually based on operations
$\cup$ and $\cap$
without inverse.
Since nature is reversible, Qunivac must be
reversible in the sense of Bennet and Fredkin.
To describe a reversible computation
we insist on a reversible set theory and  logic,
first classical and then quantum.

The only two reversible logical operations
on truth values 
are XOR  
and its negation NOTXOR.
We
arbitrarily choose XOR, as the more familiar of the two.

Boole in classical logic and Von Neumann in quantum
defined a class by an idempotent  process of selection $e^2=e$.
Selection, which is filtration,
 has no inverse. 
We define a class by a unipotent process $e^2=1$.
We may identify this new class-operation
with XOR multiplication by the class in the old sense.
The empty set is its identity element 1.
We still need an element of structure
to define the complement,
to tell a unit class from its complement,
and
to define the universal class.

We introduce a logical grade for this purpose.
This is just the modulus of the old class logic:
0 for the null set, 1 for unit sets, \... .

Our reversible classical logic (or set theory)
 is thus a graded XOR group.
We turn to the quantum logic.

When we go from the classical to the quantum logic
we turn on superposition.
The reversible quantum class or set algebra
is the Clifford algebra
generated by the unit classes or sets  \cite{FINKELSTEIN1982}.

Class algebra, reversible or not,
is the least 
interesting part of set theory for computer architects.
Since nature has a hierarchic 
structure,
Qunivac must have one,
The hierarchy-forming
operation
of classical set theory
is the power set functor $\3P: X\mapsto 2^X$. 
Finite set theory is the theory of the iterated power set functor.
We use the power set  to organize the computer,
to construct its organs.

Our quantum power set functor is
 the Clifford algebra functor Cliff: $A\mapsto {\bf 2}^A$
from graded algebras to graded algebras.

\cite{FINKELSTEIN1996} still used Grass, not 
Cliff, to form hierachy,
but that theory is unstable
and the Clifford logic is its stabilization.

We call a quantum aggregate described
by a Clifford algebra over 
the one-quantum algebra, a {\em squad}.
The qubits of a squad obey 
a real variant of Wilczek's Clifford statistics
\cite{WILCZEK1982,
NAYAK1996,WILCZEK1998}.

If the qubit of the Qunivac has
a (real) algebra $A$,
the algebra of observables of a squad of qubits,
say the entire Qunivac,
is the Clifford algebra ${\bf 2}^A$\/.
The  grade of this algebra 
counts elementary computer operations.
The elementary operations have grade 1.

The hierarchy of classical sets is built with
Peano's unitizer or successor operator $\iota z=\{z\}$,
$\iota: X\to 2^X$.
Our quantum $\iota: C\to  {\22}^C$
is a linear morphism $C\to {\22}^C$
that
transforms any element $z \in C$ into a
first-grade element $\iota  z\in C$
and  reverses the  norm 
$\|z\| :=\Re z^2$:
\begin{equation}
 (\iota z)^2 =-\|z\|\/. \|\iota z\|=-\|z\|\/.
\end{equation}
We introduce this sign reversal to generate
the indefinite metrics needed 
for relativity and gauge theory.

The operator  $\iota$ has no inverse,
since many sets are not unit sets,
but $\iota$ is reversible, in that
any unit set has a unique element.
Therefore $\iota$ has left inverse $\iota^{\6L}$:
$\iota^{\6L}\iota=\5I\5d$.
We fix $\iota^{\6L}$ uniquely
by setting it to 0 on all sets of sharp grade
other than 1.

Then we may define the reversible quantum logic 
$C$ as that represented by Cliff$(\iota)$,
the real Clifford algebra
generated by $\iota$.
Since one calls a Clifford algebra with four units  the quaternions,
we call Cliff$(\iota)$ the infinions.

This provides new 
content
to the old surmise \cite{FINKELSTEIN1969}
that  the quantum universe is a quantum computer.

We suppose that in any possible universe
a cosmical but finite number
$N$ of anticommuting
binary variables
suffices,
generating a finite-dimensional
subalgebra of the infinions
that can be called the cosmonions.

The Clifford sum provides  Qunivac 
with the famous quantum parallelism
that lets it compute so fast.

Iterated Clifford-algebra formation provides a
hierarchy-generating,
or  subprogram-forming,
function for Qunivac.

A mode of Qunivac is then represented by a  spinor
of the cosmonion algebra.

\section{Fermions}

We have programmed Qunivac for a
Dirac particle 
in a quantum space-time \cite{GALIA}.
It
respects Lorentz invariance exactly.
Its quantification
preserves  and strengthens 
the observed spin-statistics correlation,
now
giving it
 a purely algebraic origin.

On the other hand  Qunivac beats
 the standard Heisenberg uncertainty relations.
Position and momentum are now
proportional to angular momentum operators
in higher dimensions
and so is their commutator  $\eta$ \cite{GALIA}.
All three can be exactly 0
at the same time
in a singlet channel.
We expect that as in quaternion quantum field theory,
$\eta$ contracts to the  Higgs field 
in the limit $\Delta \tau\to 0, N\to \infty$
and $i\hbar$ is its effective value
in the vacuum.
The usual Heisenberg indeterminacy relations 
appear to be good
approximations only for  
for values of $\eta$ (hopefully,
the Higgs field)
close to its vacuum value.

At high energy 
$\sim \Delta\epsilon$\/,
Qunivac also violates
 the usual continuum-based locality
principle.
Elementary processes connect
events separated 
not infinitesimally as Einstein postulated but by 
a time $\Delta \tau$, the chronon.
At energies much lower than $\Delta \epsilon=\hbar /\Delta \tau$
 this would not 
show up strongly in the experimental data.

The simplest stabilization of the Dirac equation
predicts an upper bound $\Delta \epsilon/c^2$
to the mass of elementary fermions.
If we tentatively identify
this limit with  $ M(\mbox{Top quark})$,
we can estimate
$\Delta \epsilon$ and  $\Delta \tau = \Delta \epsilon/\hbar$.
The distance $c\Delta \tau$
is then  two or  three orders of magnitude smaller 
than Dehmelt's estimate
of the electron size
\cite{DEHMELT}.

It is conceivable that both Dehmelt's 
form factor size and our $\Delta \tau$ are both right.
This would, however, imply  that the electron
is quite composite,  as Dehmelt proposes.

The distance  $c\Delta \tau$ is many orders of magnitude 
greater than the Planck length.
This discrepancy does not trouble us much.
The Planck length
comes from a scenario
in which it is merely assumed 
in the absence of evidence to the contrary
that nothing limits space-time resolution
but black-hole formation.
We propose a more serious limit
arising from the structure of space-time.
In any case,
a Clifford algebraic theory like ours makes it more natural
to regard gravity as another condensation phenomenon
like the Higgs field
than a fundamental force.

\section{Fields}

Field theory begins with a partition
of variables into field and space-time.
The space-time variables are of the experimenter,
the field variables are of the system.
The set of fields field is locally an exponential $Y^X$,
where $Y$ is the field fiber and $X$ is the space-time.

To program  field theory in Qunivac
requires us to define the set exponential
$Y^X$ when the field variable space $Y$ and the space-time $X$
are both quantum.
We insist on the correspondence principle.
Our construction must have a classical limit.

To form a field in Qunivac we must first represent
the universe as a squad of events $U={\bf 2}^e$.
Then the event must reduce to a pair $e=y\otimes X$.
This factorization is a condensation or spontaneous symmetry-breaking,
reducing the orthogonal group of the event $e$
 to a product of two smaller orthogonal groups.
The field fiber at each point is a squad of filaments $Y={\bf 2}^y$.
The field universe is then $U=Y^X:={\bf 2}^{y\otimes X}$.

This is possible and easy when and apparently only when the 
field   $Y$ at each point  is a squad with Clifford algebra $Y={\bf 2}^y$.
This happens to work for spinor fields.

Since we have formulated this quantum field entirely
in Clifford algebra,
it is easy to see its classical limit.
One simply replaces $\bf 2$ by 2 throughout.

\section{Gauge}

We turn now to the bosons
and 
the gauge theory of Qunivac,
again clinging to the correspondence principle
for dear life.

Most of this section represents work
done since the Digital Perspectives symposium.

Gauge theory begins like field theory
with a division of the variables of the universe
into field and space-time variables.

The gauge group of Qunivac must be a quantum group.
We reason thus:

Loosely speaking, we recall,
a quantum group 
is a group with quantum parameters.
It is defined by an ``algebra of observables''
with two associative unital products,
the usual one for  group 
parameters
 \cite{FINKELSTEIN1996}
and an extra one for group elements.
If the parameters commute the group is classical.
If the elements commute
the group is commutative.

The gauge group of Qunivac must be a quantum group
because
 the gauge group element
depends on a point of space-time as on a parameter,
and any space-time in Qunivac is quantum.

The most recent step towards
the gauge theory of Qunivac
is to recognize that the quantum gauge group algebra
too is a Clifford algebra $G={\bf 2}^g\otimes X$.

This follows directly from the fact
that a gauge group element
is a field of Lie group elements
over a quantum space-time.

In an earlier quaternionic theory
the varying  $i\hbar$
provided the Higgs field
and reducing the gauge group.
Now  the quaternions
have spawned 
a cosmological number of Clifford elements
and the question is reopened.

A somewhat fuller exposition
is available \cite{BAUGH02}

\section{ACKNOWLEDGMENTS}
We thank 
Frank ``Tony'' Smith for algebraic information;
Frank Wilczek for his  studies of Clifford statistics;
and Shlomit Ritz Finkelstein
for discussions and help with the presentation.


\section{REFERENCES}

\begin{thebibliography}{XXX}

\bibitem{BACON} Bacon, F.   {\em Novum Organum}\/. 1620. 
Translated and edited by P.  Urbach and J. Gibson, 
Open Court Publishing Company, Peru, Illinois, 1994.  


\bibitem{BAUGH2001} Baugh, J., D. Finkelstein, A. Galiautdinov, and H.
Saller, Clifford algebra as quantum language. 
{\it J. Math. Phys.} {\bf 42}, 1489-1500 (2001)

\bibitem{BAUGH02} Baugh, J., D. Finkelstein, and A. Galiautdinov.
Elementary operation theory.
{\em International Journal of Theoretical Physics}\/. In process. 2002.


\bibitem{BERGMANN} Bergmann, P. G. The quantum state vector and physical reality.  
Chapter 3 in Bunge, M., editor, 
{\em Studies in the Foundations, Methodology, and Philosophy of Sciences.}
Vol. 2. Quantum Theory and Reality. Springer, Heidelberg (1967).

\bibitem{BLATT} Blatt, J. M. and V. F. Weisskopf, {\em Theoretical Nuclear Physics}\/.  
Wiley 1952.

\bibitem{BOHR} Bohr, N.  
Causality and Complementarity.
{\em Philosophy of Science} {\bf 4},  293-4 (1936).
Address delivered before the
Second International Congress for the Unity of Science, June, 1936.


\bibitem{DEHMELT} Dehmelt, H. G. Nobel address, 1998.

\bibitem{FINKELSTEIN1969}Finkelstein, D., Space-time code, {\it Physical Review}\/
{\bf 184}, 1261 1271 (1969); Space-time code II, {\it
Physical Review}\/ {\bf D 184}, 4-328 (1972)

\bibitem{FINKELSTEIN1982}Finkelstein, D., Quantum set theory and
Clifford algebra,  {\it International Journal of Theoretical
Physics}\/ {\bf 21}, 489-503 (1982)

\bibitem{FINKELSTEIN1996}Finkelstein, D., {\it Quantum Relativity}\/. 
Springer,
Heidelberg (1996)

\bibitem{FINKELSTEIN2001}{Finkelstein, D. R.  and A. A. Galiautdinov}.  
Cliffordons. 
{\em Journal of Mathemetical Physics} {\bf 42}, 3299-3314 (2001).

\bibitem {GALIA}Galiautdinov A.A. and D. R. Finkelstein.
 Non-local corrections to the Dirac equation. hep-th/0106273 [LANL]

\bibitem{INONU} In\"on\"u, E. and E. P. Wigner.
On the contraction of groups and their representations.
{\it Proceedings of the National Academy of Sciences} {\bf
39}(1952) 510-525. 

\bibitem{MALUS} Malus, Etienne-Louis. 
{\em M\'emoires de la Soci\'et\'e d'Arcueil} {\bf 2}, 143 (1805).
Excerpted in Magie, W. F., {\em A Source Book in Physics},
McGraw-Hill 1935.


\bibitem{NAYAK1996}Nayak, C.  and F. Wilczek. {\it Nuclear Physics}  {\bf B479} 
(1996) 529.

\bibitem{NEUMANN}Neumann, J. von (1932). 
 {\it Mathematische Grundlagen der
Quanten\-mechanik}. Berlin: Springer. Translator R. T. Beyer, 
{em Mathematical Foundations of Quantum Mechanics}\/. Princeton, 1955.

\bibitem{SEGAL}Segal, I. E.  A class of operator algebras which
are determined by groups. {\it Duke Mathematics Journal} {\bf 18} (1951) 221.

\bibitem{WHEELER}Wheeler, J. A. 
The elementary quantum act as higgledy-piggledy building mechanism. 
In Castell, L., M. Drieschner, and C. F. v. Weizsa\"cker (editors),
{\em Quantum Theory and the Structure of Time and Space}\/. IV. 
Hanser, Munich (1973). Pages 27-30.

\bibitem{WILCZEK1982}Wilczek, F. and A. Zee (1982) Families from spinors.
{\em Physical Review} {\bf D25}, 553

\bibitem{WILCZEK1998}Wilczek, F. Projective statistics and spinors in
 Hilbert space, hep-th/9806228 [LANL]

\end{thebibliography}
\end{document}